\documentclass[amsthm]{amsart}

\usepackage{mylogic,amsmath,amssymb}

\newcommand{\Z}{\mathbb{Z}}

\begin{document}


\title[Q.E. for the reals with a predicate for the powers of two]{Quantifier elimination for the reals with\\
  a predicate for the powers of two}

\author{Jeremy Avigad and Yimu Yin}

\date{May 9, 2006}

\thanks{To appear in \emph{Theoretical Computer Science}.}



\maketitle

\begin{abstract}
  In \cite{van:den:dries:85}, van den Dries showed that the theory of
  the reals with a predicate for the integer powers of two admits
  quantifier elimination in an expanded language, and is hence
  decidable. He gave a model-theoretic argument, which provides no
  apparent bounds on the complexity of a decision procedure. We
  provide a syntactic argument that yields a procedure that is
  primitive recursive, although not elementary. In particular, we show
  that it is possible to eliminate a single block of existential
  quantifiers in time $2^0_{O(n)}$, where $n$ is the length of the
  input formula and $2_k^x$ denotes $k$-fold iterated exponentiation.
\end{abstract}




\section{Introduction}
\label{introduction:section}

Consider the theory of real closed fields, in a language with $0, 1,
+, -, \times, <$. Extend the language with a predicate $A$ which, in
the intended interpretation, holds of the powers of two, $2^\Z$.
Adopting the obvious conventions and abbreviations, extend the theory
by adding the following axioms:
\begin{itemize}
\item $\fa x (A(x) \limplies x > 0)$
\item $\fa {x,y} (A(x) \limplies (A(y) \liff A(xy)))$
\item $A(2) \land \fa x (1 < x < 2 \limplies \lnot A(x))$
\item $\fa x (x > 0 \limplies \ex y (A(y) \land y \leq x < 2y))$
\end{itemize}
The first two imply that the $A$ picks out a multiplicative subgroup
of the positive elements. In \cite{van:den:dries:85}, van den Dries
showed that the resulting theory admits quantifier elimination in an
expanded language. As a result, it is complete and decidable, and, in
particular, axiomatizes the real numbers with a predicate for the
powers of two. 

The theory we have just described includes not only the theory of
real closed fields, but also, via an interpretation of integers as
exponents, Presburger arithmetic. Thus, van den Dries's result is
particularly interesting in that it subsumes two of the most important
decidability results of the twentieth century. In recent years, this
result has been extended in various directions (see, for example,
\cite{friedman:miller:05,van:den:dries:gunaydin:unp}). 

To establish quantifier-elimination, van den Dries gave a
model-theoretic argument. The proof does not provide an explicit
procedure, nor does it provide a bound on the length of the resulting
formula. Here, we present a proof that makes use of nested calls to a
quantifier-elimination procedure for real closed fields, yielding a
procedure that is primitive recursive but not elementary. In
particular, it requires time $2_{O(n)}^0$ to eliminate a single block
of existential quantifiers, or even a single existential quantifier,
where $n$ is the length of the input formula and $2_k^0$ denotes
a stack of $k$ exponents. Thus, the best bound we can give on the time
complexity of the full quantifier-elimination procedure involves
$O(n)$ iterates of the stack-of-twos function. We leave it as an open
question as to whether one can avoid such nesting and, say, obtain
elementary bounds for the elimination of a single existential
quantifier.

In Section~\ref{first:section}, we describe the extension of the
theory above that admits elimination of quantifiers. Our method of
eliminating an existential quantifier proceeds in two steps: first, we
eliminate that quantifier in favor of a multiple existential
quantifiers over powers of two (the number of which is bounded by the
length of the original formula); then we successively eliminate each
of these. The first step is described in Section~\ref{first:section}.
In Section~\ref{reasoning:section}, we prove a number of lemmas that
fill out the relationship between the powers of two and the underlying
model of real closed fields in a model of the relevant theory; this
contains the bulk of the syntactic and algebraic work. In
Section~\ref{second:section}, we use these results to carry out the
second step. Finally, in Section~\ref{complexity:section}, we show
that our procedure satisfies the complexity bounds indicated above.


We are grateful to Chris Miller for bringing van den Dries's result to
our attention, and for raising the issue of finding an explicit
elimination procedure. We are also grateful to the anonymous referees
for comments and corrections.


\section{The first step}
\label{first:section}

Expand the language of real closed fields to include a unary function
$\lambda$ and a unary predicate $D_n$ for each natural number $n \geq
1$. Let $T$ be the theory given by the axioms above together with the
following:
\begin{itemize}
\item $D_n(x) \liff \ex y (A(y) \land y^n = x)$
\item $\fa x (x \leq 0 \limplies \lambda(x) = 0)$
\item $\fa x (x > 0 \limplies A(\lambda(x)) \land \lambda(x) \leq x < 2
  \lambda(x))$
\end{itemize}
In the standard interpretation, $\lambda$ maps negative real numbers
to $0$ and rounds positive reals down to the nearest power of two, and
$D_n$ holds of numbers of the form $2^i$ where $i$ is an integer
divisible by $n$.\footnote{For parsimony, $0$ can be defined as $1 -
  1$ and $A(x)$ by $x > 0 \land \lambda(x) = x$. In the next section,
  we will see that the division symbol is another inessential addition
  to the language. But in contrast to q.e.~for real closed fields, one
  can't eliminate $-$ in terms of $+$; for example, the
  quantifier-free formula $A(x-y)$, if replaced by $\ex z (z + y = x
  \land A(z))$, would have no quantifier-free equivalent.} Note that
$A$ and $D_1$ are equivalent; we will treat them as the same symbol
and use the two notations interchangeably.

Our goal is to prove the following:
\begin{theorem}
\label{main:thm}
$T$ admits quantifier-elimination.
\end{theorem}
This is Theorem II of \cite{van:den:dries:85}. Henceforth, by
``formula,'' we mean ``formula in the language of $T$.'' We will use
$\vec x$ to denote a sequence of variables $x_0, x_1, \ldots,
x_{k-1}$, and we will use notation like $A(\vec x)$ to denote $A(x_0)
\land A(x_1) \ldots \land A(x_{k-1})$.

To eliminate quantifiers from any formula it suffices to be able to
eliminate a single existential quantifier, i.e.~transform a formula
$\ex x \ph$, where $\ph$ is quantifier-free, to an equivalent
quantifier-free formula. Since $\ex x (\ph \lor \psi)$ is equivalent
to $\ex x \ph \lor \ex x \psi$, we can always factor existential
quantifiers through a disjunction. In particular, since any
quantifier-free formula can be put in disjunctive normal form, it
suffices to eliminate existential quantifiers from conjunctions of
atomic formulas and their negations.  Also, since $\ex x (\ph \land
\psi)$ is equivalent to $\ex x \ph \land \psi$ when $x$ is not free in
$\psi$, we can factor out any formulas that do not involve $x$.
Furthermore, whenever we can prove $\fa x (\theta \lor \eta)$, $\ex x
\ph$ is equivalent to $\ex x (\ph \land \theta) \lor \ex x (\ph \land
\eta)$; so we can ``split across cases'' as necessary. We will use
all of these facts freely below.

In \cite{van:den:dries:85}, van den Dries established quantifier
elimination by demonstrating the following two facts:
\begin{enumerate}
\item Every model of $T^\forall$, the universal fragment of $T$, has a
  ``$T$-closure''; in other words, every model $\mdl M$ of $T^\forall$
  can be extended to a model of $T$ which can be embedded, over $\mdl
  M$, into any other $T$-extension of $\mdl M$.  
\item If $\mdl M$ is a proper substructure of $\mdl N$ and both are
  models of $T$, there is some $b \in \mdl N - \mdl M$ such that $\mdl
  M(b)$, the model of $T^\forall$ generated by $\mdl M \cup \{ b \}$,
  can be embedded into an elementary extension of $\mdl M$.
\end{enumerate}
The novelty of this test, as compared to more common ones (see e.g.\ 
\cite{shoenfield:01}), lies in the prover's right to choose an
appropriate $b$ in the second clause (see also the discussion in
\cite{van:den:dries:88}). This clause implies that any existential
formula with parameters from $\mdl M$ that is true in the $T$-closure
of $\mdl M(b)$ is true in $\mdl M$; the test works because this clause
can be iterated in a countable model to obtain a sequence of
$T$-extensions $\mdl M = \mdl M_0 \subseteq \mdl M_1 \subseteq \mdl
M_2 \ldots \subseteq \mdl N$ that eventually picks up every element of
$\mdl N$, so any existential formula with parameters from $\mdl M$
true in $\mdl N$ is true in $\mdl M$. On the syntactic side, this
iteration translates to the simple observation that to eliminate a
single existential quantifier from an otherwise quantifier-free
formula, it suffices to eliminate additional existential quantifiers
from an equivalent existential formula. Thus, our effective proof is
based on the following two lemmas:

\begin{lemma}
\label{main:lemma:one}
  Every formula of the form $\ex w \psi$, with $\psi$ quantifier-free,
  is equivalent to a disjunction of formulas of the form $\ex {\vec x}
  (A(\vec x) \land \ph)$, with $\ph$ quantifier-free.
\end{lemma}

\begin{lemma}
\label{main:lemma:two}
Every formula of the form $\ex x (A(x) \land \ph)$, with $\ph$
quantifier-free, is equivalent to a formula that is quantifier-free.
\end{lemma}

The remainder of this section is devoted to proving the first of these
two lemmas. The next lemma explains why the new existentially
quantified variables are helpful.

\begin{lemma}
\label{nice:lemma}
Every existential formula is equivalent, in $T$, to an existential
formula in which $\lambda$ does not occur and the predicates $D_i$ are
applied only to variables.
\end{lemma}

\begin{proof}
  First, replace $\ldots D_i(t) \ldots$ by $\ex z (z = t \land \ldots
  D_i(z) \ldots)$. Then, iteratively simplify terms
  involving $\lambda$, noting that $\psi(\lambda(t))$ is equivalent to
\[
(t \leq 0 \land \psi(0)) \lor \ex z (A(z) \land z \leq t < 2z \land
\psi(z)),
\]
and that the existential quantifier can be brought to the front.
\end{proof}

Thus to prove Lemma~\ref{main:lemma:one}, we are reduced to showing
that when $\psi$ is quantifier-free, $\lambda$ does not occur in
$\psi$, and the predicates $D_i$ occurring in $\psi$ are applied only
to variables, the formula $\ex {\vec x} \psi$ is equivalent to one of
the form $\ex {\vec x} (A(\vec x) \land \ph)$, where $\ph$ is
quantifier-free. In general, $\ex x \theta(x)$ is equivalent to
\[
\ex {x > 0} \theta(x) \lor \theta(0) \lor \ex {x > 0} \theta(-x).
\]
Moreover, assuming $x > 0$, any subformula of the form $D_i(-x)$ is
equivalent to falsity. So, across a disjunction, we are reduced to
proving the claim for formulas of the form $\ex {\vec x > 0} \psi(\vec
x)$, where $\psi$ satisfies the criteria above.

In $T$ we can factor out the greatest power of two from any
positive $x$, i.e.\ we can prove
\[
x > 0 \liff \ex y \ex z (A(y) \land 1 \leq z < 2 \land x = y z).
\]
Since we have $1 \leq z < 2 \liff (z = 1 \lor 1 < z < 2)$,
we can transform our formula into a disjunction of formulas of the
form
\[
\ex {\vec y, \vec z} (A(\vec y) \land 1 < \vec z < 2 \land \psi)
\]
where $\psi$ once again meets the criteria above, except that the
predicates $D_i$ are applied to expressions of the form $yz$. When $1
< z < 2$, each $D_i(yz)$ is false, so we can rewrite the formula above
as
\[
\ex {\vec y} (A(\vec y) \land \theta \land \ex {\vec z}
\eta)
\]
where $\theta$ is a conjunction of predicates of the form $D_n(y)$ and
negations of such, and $\ex {\vec z} \eta$ is in the language of real
closed fields. We can therefore replace $\ex {\vec z} \eta$ by a
quantifier-free formula, using any q.e.\ procedure for real closed fields.


\section{Reasoning about powers of two}
\label{reasoning:section}

Our goal in this section is to establish some general relationships
between the powers of two in a model of our theory, $T$, and the
underlying real closed field.

\begin{definition}
Let $\varphi$ be a quantifier-free formula. We say $\varphi$ is
\emph{simple in $x$} if the following hold:
\begin{enumerate}
\item every equality or inequality occurring in $\ph$ is either of the
  form $p(x) = 0$ or $q(x) > 0$, where $p(x)$, $q(x)$ are polynomials
  in $x$; that is, they are of the form $\sum_{i \leq n} s_i x^i$ where
  each $s_i$ is a term that does not involve $x$.
\item for every atomic formula $D_n(t)$ occurring in $\varphi$,
  either $t$ does not contain $x$ or $t$ is of the form
  $2^r x$ for some integer $r$ such that $0 \leq r < n$.
\end{enumerate}
\end{definition}

The main goal of this section is to prove the following proposition:

\begin{proposition}
\label{main:prop}
Let $\varphi$ be any quantifier-free formula. Then there is a
quantifier-free formula $\varphi'$ such that $\varphi'$ is simple in
$x$ and $T$ proves $A(x) \limplies (\varphi \leftrightarrow
\varphi')$.
\end{proposition}

In semantic terms, this says the following: let $\mdl N$ be any model
of $T$, let $\mdl M \subseteq \mdl N$ be a model of $T^\forall$, and
let $x$ be a power of two in $\mdl N$. Then the structure of $\mdl
M(x)$ is completely determined by the structure of $\mdl M$, the
structure of $\mdl M(x)$ as an ordered ring, and the divisibility
properties of the exponent of $x$.

First, we need to note some easy facts about $\lambda$ and the
predicates $D_i$. 

\begin{lemma}
\label{finite:disjunction:lemma}
For any $n$, $T$ proves
\[
0 < u < x \leq 2^n u \land A(x) \limplies (x = 2
\lambda(u) \lor \ldots \lor x = 2^n \lambda(u)).
\]
\end{lemma}

\begin{lemma}
\label{finite:disjunction:lemma:two}
For any $n$, $T$ proves
\[
A(x) \limplies D_n(x) \lor D_n(2x) \lor \ldots \lor D_n(2^{n-1} x).
\]
\end{lemma}

Although we have not included the division symbol in the language of
$T$, we can define the function $r / s$ by making $x / y = z$
equivalent to $x = y z \lor (y = 0 \land z = 0)$. In the proof of
Proposition~\ref{main:prop}, it will be useful to act as though the
division symbol is part of the language. The next few lemmas show that
if $\theta$ is any quantifier-free formula in the expanded language
with division, there is a quantifier-free formula $\theta'$ in the
language without division such that $T \proves \theta \liff \theta'$.

\begin{lemma} 
\label{elim:one:div:lambda:lemma}
From the hypotheses $0 < x$ and $0 < y$, $T$ proves
\[
x \lambda(y) < y \lambda(x) \limplies \lambda(x/y) = \lambda(x) / 2
\lambda(y)
\]
and
\[
x \lambda(y) \geq y \lambda(x) \limplies \lambda(x/y) = \lambda(x) /
\lambda(y).
\]
\end{lemma}

\begin{proof}
  An easy calculation using the axioms for $\lambda$ shows that if $x
  / y < \lambda(x) / \lambda(y)$, then $\lambda(x / y) = \lambda(x) /
  2 \lambda(y)$; and otherwise, $\lambda(x / y) = \lambda(x) /
  \lambda(y)$.
\end{proof}

\begin{lemma}
\label{elim:div:lambda:lemma}
If $\theta$ is any quantifier-free formula involving the division
symbol, there is a quantifier-free formula $\theta'$ in which the
division symbol does not occur in the scope of $\lambda$, such that $T
\proves \theta \liff \theta'$.
\end{lemma}

\newcommand{\lamdiv}[1]{\Lambda^{\mathord{\div}}(#1)}

\begin{proof}
  This can be done by iterating the previous lemma. To measure the
  nesting of $\lambda$'s and division symbols, we define the
  ``$\lambda$-depth of the division symbol in $t$,'' $\lamdiv{t}$,
  recursively, as follows:
\begin{enumerate}
\item $\lamdiv{t} = 0$ if the division symbol does not occur in the
  scope of $\lambda$ in $t$;
\item if $t$ is $t_1 + t_2$, $t_1 - t_2$, $t_1 \times t_2$, or $t_1 /
  t_2$, then $\lamdiv{t}=\max \{\lamdiv{t_1}, \lamdiv{t_2} \}$;
\item assuming the division symbol occurs in $t$, $\lamdiv{\lambda(t)}
  = \lamdiv{t} + 1$.
\end{enumerate}
The previous lemma shows that, using a case disjunction over the
possibilities for the signs of the numerator and denominator, we can
eliminate one term $t$ such that the $\lambda$-depth of the division
symbol in $t$ is maximal, in favor of terms in which the
$\lambda$-depth of the division symbol is smaller.
Lemma~\ref{elim:div:lambda:lemma} follows, by a primary induction on
this maximal depth, and a secondary induction on the number of terms
of this depth.
\end{proof}

\begin{lemma}
\label{elim:one:div:prop}
$T \proves A(x) \land A(y) \limplies (D_n(x/y) \liff \bigvee_{i<n}
(D_n(2^i x) \land D_n(2^i y)))$.
\end{lemma}

\begin{proof}
  The right-to-left direction is easy: if $z^n = 2^i x$ and $w^n = 2^i
  y$ then $(z/w)^n = x / y$. Proving the other direction is not much
  more difficult, using Lemma~\ref{finite:disjunction:lemma:two}.
\end{proof}

\begin{proposition}
\label{elim:div:prop}
Let $\theta$ be any quantifier-free formula involving division. Then
there is a quantifier-free formula $\theta'$ that does not involve
division, such that $T \proves \theta \liff \theta'$.
\end{proposition}

\begin{proof}
Using Lemma~\ref{elim:div:lambda:lemma}, we can assume that division
does not occur in the scope of any $\lambda$ in $\theta$. So each
atomic formula $D_n(t)$ can be put in the form $D_n(r/s)$, where the
division symbol does not occur in $r$ and $s$. Across a case disjunct,
we can assume $r$ and $s$ are positive. Then $D_n(r/s)$ is
equivalent to
\[
\lambda(r/s) = r / s \land D_n(\lambda(r/s)).
\]
Using Lemma~\ref{elim:one:div:lambda:lemma}, we can replace
$\lambda(r/s)$ by either $\lambda(r) / \lambda(s)$ or $\lambda(r) / 2
\lambda(s)$. Then using Lemma~\ref{elim:one:div:prop} we can replace
$D_n(\lambda(r) / \lambda(s))$ or $D_n(\lambda(r) / 2\lambda(s))$ by a
disjunction in which the division symbol does not occur.

Once all divisibility symbols are removed from the $\lambda$'s and
$D_n$'s, we can clear division from the remaining equalities and
inequalities by multiplying through.
\end{proof}

It therefore suffices to prove Proposition~\ref{main:prop} where
$\ph'$ is a quantifier-free formula in the expanded language with the
division symbol. The next few lemmas, then, make use of this expanded
language.

\begin{lemma}
\label{algebra:one:lemma}
Let $p(x)$ be the term $\sum_{i \leq n}a_i x^i$. Then there is a
sequence of quantifier-free formulas $\theta_0, \ldots, \theta_{m-1}$
such that $T$ proves
\[
A(x) \wedge p(x) > 0 \limplies  \bigvee_{k < m} \theta_k, 
\]
where each $\theta_k$ is of one of the following forms:
\begin{itemize}
 \item $\lambda(p(x)) = 2^r\lambda(a_i)x^i$ for some $-1 \leq r \leq n$,
 \item $x^e = \frac{2^r\lambda(a_i)}{\lambda(- a_j)}$ or $x^e =
   \frac{2^r\lambda(- a_j)}{\lambda(a_i)}$, for some $e, i, j$, and
   $r$ such that $1 \leq e \leq n$, $0 \leq i,j \leq n$, and $-(n+1)
   \leq r \leq (n+1)$.
\end{itemize}
\end{lemma}

\begin{proof}
  Argue in $T$. Using a disjunction on all possible cases, we can
  write $p(x)$ as $a_i x^i + a_j x^j + \hat p(x)$, where $a_i x^i$ is
  the largest summand and $a_jx^j$ the least summand. Note that we
  have $a_ix^i > 0$, $p(x) \leq (n + 1) a_i x^i$, and
\[
  p(x) - a_i x^i = a_j x^j + \hat p(x) \geq n a_j x^j.
\] 
We now distinguish between two cases, depending on whether $p(x)$ is
roughly the same size as $a_i x^i$ or sufficiently smaller.

In the first case, suppose we have $p(x) \geq (a_i x^i) / 2$. This
means we have
\[
(a_i / 2) x^i \leq p(x) \leq (n + 1) a_i x^i \leq 2^n a_i x^i
\]
which yields
\[
(\lambda(a_i) / 2) x^i \leq \lambda(p(x)) \leq 2^n \lambda(a_i) x^i.
\]
This yields a disjunction of clauses of the first type, by
Lemma~\ref{finite:disjunction:lemma}. 

In the second case, we have $p(x) < (a_i x^i) / 2$ and $i \neq j$.
This means that $a_j x^j$ must be negative and roughly comparable to
$a_i x^i$ in absolute value. That is, we have $a_j < 0$ and
\[
(a_i / 2) x^i < a_i x^i - p(x) \leq - n a_j x^j,
\]
and so
\[
(a_i / (- a_j)) x^{i-j} \leq 2 n \leq 2^n.
\]
Also, $p(x) > 0$ implies $na_i x^i \geq -a_j x^j$, which yields
\[
0 < 2^{-n} < 1/n \leq (a_i / (- a_j)) x^{i-j}.
\]
Combining these, we have $2^{-n} < (a_i / (- a_j)) x^{i-j} \leq
2^n$. Using Lemma~\ref{elim:one:div:lambda:lemma} and
Lemma~\ref{finite:disjunction:lemma} we get a disjunction of clauses
of the second type.
\end{proof}

\begin{lemma}
\label{algebra:two:lemma}
In Lemma~\ref{algebra:one:lemma}, if the assumption is changed to
$A(x) \wedge p(x) = 0$, then in the conclusion we can assume that each
$\theta_k$ is of the second form.
\end{lemma}

\begin{proof}
This is exactly as in the second case of the previous proof.
\end{proof}

\begin{lemma}
\label{algebra:three:lemma}
In the conclusion of Lemma~\ref{algebra:one:lemma}, we may demand that
each $\theta_k$ is of the form $\lambda(p(x)) = sx^i$ for some $0 \leq
i \leq n$ and some term $s$ that does not contain $x$.
\end{lemma}

\begin{proof}
The proof is by induction on the degree of $x$ in $p(x)$. The lemma
is trivial if the degree of $x$ in $p(x)$ is 0.

Now assume that the degree of $x$ in $p(x)$ is $n$ and the lemma holds
whenever the degree is less than $n$. By
Lemma~\ref{algebra:one:lemma}, $T$ proves a disjunction $\bigvee
\sigma_l$, with $\sigma_l$ of one of those two forms. Each $\sigma_l$
of the first form there is already as required. For each $\sigma_l$ of
the second form, consider a new term $\hat p(x)$, which is obtained by
substituting the right-hand side of $\sigma_l$ for $x^e$ in $p(x)$.
Notice that the degree of $x$ in $\hat p(x)$ is less than $n$, and
clearly $T$ proves $p(x) = \hat p(x) \land \hat p(x) > 0$. By the
inductive hypothesis we may replace $\sigma_l$ in $\bigvee \sigma_l$
by a disjunction $\bigvee \theta_k$ which is of the required form.
\end{proof}

As was the case with the division symbol, we will iterately
``squeeze'' $x$'s out from within the $\lambda$ symbols. Thus we
introduce the following definitions:

\begin{definition}
\label{lambda:depth:def}
Let $t$ be a term. Define the \emph{$\lambda$-depth of $x$ in $t$},
$\Lambda(x,t)$, recursively, as follows:
\begin{enumerate}
      \item $\Lambda(x,t)=0$ if $x$ is not in the scope of any $\lambda$;
      \item if $t$ is $t_1 + t_2$, $t_1 - t_2$, $t_1 \times t_2$, or
        $t_1 / t_2$, then $\Lambda(x,t)=\max \{\Lambda(x,t_1),
          \Lambda(x,t_2)\}$;
      \item if $t$ is $\lambda(t_1)$ and $t_1$ contains $x$, then
        $\Lambda(x,t)=\Lambda(x,t_1)+1$.
\end{enumerate}
\end{definition}

\begin{definition}
Let $\varphi$ be a formula. Define the \emph{$\lambda$-depth of $x$ in
$\varphi$} by
\[\Lambda(x,\varphi)=\max\{\Lambda(x,t): 
t \text{ is a term that contains } x \text{ and occurs in }
\varphi\}.\]
\end{definition}

\begin{lemma}
\label{algebra:four:lemma}
Let $\varphi$ be any quantifier-free formula. Then there is a
quantifier-free formula $\ph'$ such that $T \proves A(x) \limplies (\ph
\liff \ph')$, and $\Lambda(x,\ph') = 0$.
\end{lemma}

\begin{proof}
The proof is by induction on the $\lambda$-depth of $x$ in
$\varphi$. The lemma is trivial if $\Lambda(x, \varphi) = 0$.

Assume $\Lambda(x, \varphi) = n > 0$ and the lemma holds for every
quantifier-free formula $\psi$ if $\Lambda(x, \psi) < n$. Let
$\lambda(p_0)$, $\ldots$, $\lambda(p_{m-1})$ be all the different
terms in $\varphi$ with $\Lambda(x, p_i) = 0$ for all $i < m$.
Across a case disjunction we can assume $p_i > 0$ for all $i < m$,
since otherwise we can replace $\lambda(p_i)$ by $0$.
By Lemma~\ref{elim:div:lambda:lemma}, we may assume
that each $p_i$ is a polynomial in $x$. 
By Lemma~\ref{algebra:three:lemma}, $T$
proves $\varphi \leftrightarrow \bigvee (\tau_l \wedge \sigma_l)$,
where each $\tau_l$ is of the form $\bigwedge_{i < m}
\lambda(p_i(x)) = s_ix^{j_{i}}$, and each $\sigma_l$ is obtained by
substituting $s_ix^{j_{i}}$ for $\lambda(p_i)$ in $\varphi$. Clearly
$T$ proves 
\[
A(x) \limplies (\lambda(p_i(x)) = s_ix^{j_{i}} \leftrightarrow A(s_i)
\wedge s_ix^{j_{i}} \leq p_i(x) < 2s_ix^{j_{i}}). 
\]
Now since $\Lambda(x, \sigma_l) < n$, we may apply the inductive
hypothesis to each $\sigma_l$ and the lemma is proved.
\end{proof}

\begin{lemma}
\label{algebra:five:lemma}
Let $p$ be a term such that $\Lambda(x, p) = 0$. Then for any $n$
there is a sequence of terms $p_k$ such that
\begin{itemize}
\item $T$ proves $A(x) \land p > 0 \limplies (D_n(p) \leftrightarrow
  \bigvee (p = p_k \wedge D_n(p_k)))$,
 \item each $p_k$ is of the form $sx^i$, where $s$ is a term
 that does not contain $x$.
\end{itemize}
\end{lemma}

\begin{proof}
  Using Lemma~\ref{elim:one:div:prop}, we can assume that $p$ is a
  polynomial in $x$. We can replace $D_n(p)$ by $p = \lambda(p) \land
  D_n(\lambda(p))$, and then by Lemma~\ref{algebra:three:lemma},
  across a disjunction we may replace $\lambda(p)$ in each disjunct by
  a term of the form $sx^i$, where $s$ does not contain $x$.  (Note
  that here no formulas like the $\tau_l$'s in the previous lemma are
  needed.)
\end{proof}

\begin{lemma}
\label{algebra:six:lemma}
Let $s$ be a term that does not contain $x$. Then for
any $n$, $i$ there is a sequence of formulas $\theta_k$ such that
$T$ proves 
\[
A(x) \limplies (D_n(sx^i) \leftrightarrow \bigvee \theta_k), 
\]
and each $\theta_k$ is of the form $D_n(2^ws) \wedge D_n(2^rx)$ for
some $0 \leq w, r < n$.
\end{lemma}

\begin{proof}
Since for each $n$, from the assumption $A(x)$, $T$ proves
$\bigvee_{j<n}D_n(2^j x)$, it is straightforward to see that
$D_n(sx^i)$ is equivalent to a disjunction each of whose disjuncts
is of the specified form.
\end{proof}

We are finally ready to prove Proposition~\ref{main:prop}.

\begin{proof}
  Given $\ph$, first use Lemma~\ref{algebra:four:lemma} to eliminate
  $x$ from the scope of any $\lambda$. Then use
  Lemma~\ref{algebra:five:lemma} to ensure the atomic formulas
  involving $D_n$ are in the form $D_n(sx^i)$, where $s$ does not
  involve $x$. (This will require splitting across cases depending on
  whether $p > 0$ or $p \leq 0$; in the latter case, $D_n(p)$ is
  equivalent to $\bot$.) Finally, use Lemma~\ref{algebra:six:lemma} to
  ensure that all the atomic formulas involving $D_n$ are in the
  required form.
\end{proof}

We close with some consideration about the predicates $D_n$ which are
analogous to considerations that arise in the context of
quantifier-elimination for Presburger arithmetic. Remember that when
$n$ is a positive integer and $s$ is a non-negative integer, $D_n(2^s
x)$ asserts, in the intended interpretation, that $x$ is equal to $2^t$
for some integer $t$, and $n$ divides $s + t$; in other words, the
exponent of $x$ is congruent to $-s$ modulo $n$. Let $\theta$ be any
boolean combination of predicates of the form $D_n(2^s x)$, and let
$M$ be the least common multiple of these various $n$. Then in $T$ one
can show that there is an $x$ satisfying $\theta$ if and only if for
any $w$ satisfying $A(w)$ we have
\[
\theta(w) \lor \theta(2w) \lor \theta(4w) \lor \ldots \lor
\theta(2^{M-1}w),
\]
and, in particular, if and only if
\[
\theta(1) \lor \theta(2) \lor \theta(4) \lor \ldots \lor \theta(2^{M-1}).
\]
Moreover, $T$ can decide the truth or falsity of this last
sentence. So we have:

\begin{lemma}
\label{theta:m:lemma}
  With $\theta$ and $M$ as above, either $T$ proves $\fa x \lnot
  \theta$, or it proves
\[
\fa u (0 < u \limplies \ex x (u \leq x < 2^M u \land \theta)).
\]
\end{lemma}


\section{Eliminating a quantifier over powers of two}
\label{second:section}

We are now ready to prove Lemma~\ref{main:lemma:two}, which asserts
that every formula of the form $\ex x (A(x) \land \ph)$, with $\ph$
quantifier-free, is equivalent to a formula that is quantifier-free.
By Proposition~\ref{main:prop}, we can assume that $\ph$ is simple,
which is to say, $x$ does not occur in the scope of any $\lambda$ and
all divisibility assertions involving $x$ are of the form $D_n(2^r
x)$. Put $\ph$ in disjunctive normal form, replace negated equalities
$s \neq t$ by $s < t \lor t < s$, and replace negated inequalities $s
\not < t$ by $t < s \lor t = s$. Rewrite equalities and inequalities
so that they are of the form $p(x) = 0$ and $q(x) > 0$, where $p(x)$
and $q(x)$ are polynomials in $x$. Factoring existential quantifiers
through disjunctions and getting rid of atomic formulas that do not
depend on $x$, we are reduced to eliminating quantifiers of the form
$\ex x (A(x) \land \ph)$ where $\ph$ is a conjunction of formulas of
the following types:
\begin{itemize}
\item $p(x) = 0$, where $p$ is a polynomial, 
\item $q(x) > 0$, where $q$ is a polynomial, 
\item $D_n(2^r x)$, where $0 \leq r < n$, or
\item $\lnot D_n(2^r x)$, where $0 \leq r < n$.
\end{itemize}
Splitting across a disjunction, we can assume that in a conjunct of
the form $p(x) = 0$, not all the coefficients are zero. By
Lemma~\ref{algebra:two:lemma}, we can assume that one of the conjuncts
is of the form $x^e = s$, where $x$ does not occur in $s$. In that
case, each conjunct $D_n(2^r x)$ is equivalent to $D_{ne}(2^{re} x^e)$
and hence $D_{ne}(2^{re} s)$ (and $A(x)$, in particular, is equivalent
to $D_e(s)$). But now $x$ no longer occurs in these formulas, and so
they can be brought outside the scope of the existential quantifier.
The resulting existential formula is then essentially in the language
of real closed fields. By this last phrase we mean that it is of the
form $\ex x \alpha(x, t_0,\ldots,t_{k-1})$, where $\alpha(x,
y_0,\ldots,y_{k-1})$ is in the language of real closed fields.
Treating the terms $t_0, \ldots, t_{k-1}$ in the expanded language as
parameters, we can therefore replace it by an equivalent
quantifier-free formula using any q.e.\ procedure for real closed
fields.

We are thus reduced to eliminating an existential quantifier of the
form
\begin{equation}
\label{main:eq}
\ex x (\bigwedge q_i(x) > 0 \land \theta(x))
\end{equation}
where $\theta$ is a conjunction of formulas of the form $D_n(2^r x)$
and negations of such that includes at least the formula $A(x)$. By
Lemma~\ref{theta:m:lemma}, either $T$ proves that $\theta$ is false for
every $x$, or there is a natural number $M$ such that $T$ proves that
for any $u > 0$, that $\theta$ is satisfied by some $x$ in the
interval $[u,2^Mu]$. In the first case, $T$ proves that
formula~(\ref{main:eq}) is false. So we only have to worry about the
second case. Fix such an $M$ for the remainder of the discussion.

Arguing in $T$, suppose formula~(\ref{main:eq}) holds. There are two
possibilities: either there is a ``large'' interval on which
$\bigwedge q_i(x) > 0$, that is, an interval of the form $[u,2^M u]$;
or there is an $x$ satisfying $A(x) \land \bigwedge q_i(x) > 0 \land
\theta$, but it is trapped between a $u$ and a $v$ with $q_i(u) = 0$
for some $i$, $q_j(v) = 0$ for some $j$, and $v < 2^M u$. Thus
formula~(\ref{main:eq}) is equivalent to a disjunction of the formula
\[
\ex {u > 0} \fa x (u \leq x \leq 2^M u \limplies \bigwedge q_i(x) > 0)
\]
and the formulas
\[
\ex {u > 0} (q_j(u) = 0 \land \ex x (u < x \leq 2^M u \land
\bigwedge q_i(x) > 0 \land \theta(x))
\]
for the various $j$. To see this, note that if formula~(\ref{main:eq})
holds, then by the previous discussion one of these formulas holds;
and conversely, each of these formulas implies (\ref{main:eq}).

The first of these formulas is essentially in the language of real
closed fields, so these quantifiers can be eliminated. The second
formula is equivalent to
\begin{multline*}
  \ex {u_1, u_2} (A(u_1) \land 1 \leq u_2 < 2 \land q_j(u_1 u_2) = 0 \mathop{\land} \\
  \ex x (u_1 < x \leq 2^M u_1 \land \bigwedge q_i(x) > 0 \land
  \theta(x)).
\end{multline*}
In this case, we can replace the inner existential quantifier over $x$
by a disjunction, so that the entire formula is equivalent to a
disjunction of formulas of the form
\[
\ex {u_1, u_2} (A(u_1) \land 1 \leq u_2 < 2 \land q_j(u_1 u_2) = 0 \land 
\bigwedge \hat q_i(u_1) > 0 \land \hat \theta(u_1)),
\]
where each $\hat q_i(u_1)$ is $q_i(2^r u_1)$ for some $r$ such that $1
\leq r \leq M$, and similarly for $\hat \theta(u_1)$. In particular,
$\hat \theta(u_1)$ is a conjunction of formulas of the form $D_i(2^r
u_1)$, and their negations.

Think of $q_j(u_1 u_2)$ as a polynomial in $u_1$ with coefficients of
the form $s u_2^n$, where $s$ does not involve $u_1$ or $u_2$. By
Lemma~\ref{algebra:two:lemma}, across a disjunction we may add a
clause of the form $u_1^e = 2^r \lambda(s u_2^n) / \lambda(t u_2^m)$.
Splitting on cases of the form $2^l \leq u_2^h < 2^{l+1}$ we can
simplify each of these to an expression of the form $u_1^e = 2^k
\lambda(s) / \lambda(t)$ for some integer $k$. By
Lemma~\ref{theta:m:lemma}, $A(u_1) \land \hat \theta(u_1)$ is
equivalent to a formula $\bar \theta$ which now involves neither $u_1$
nor $u_2$, and hence can be brought outside the existential quantifier.
We are thus reduced to eliminating quantifiers from a formula of the
form
\begin{multline*}
\ex {u_1, u_2} (1 \leq u_2 < 2 \land u_1^e = 2^k \lambda(s) /
\lambda(t) \land 2^l \leq u_2^h < 2^{l+1} \land \; \\
q_j(u_1 u_2) = 0 \land
\bigwedge \hat q_i(u_1) = 0).
\end{multline*}
We can eliminate these quantifiers using a q.e.~procedure for real
closed fields. This completes the proof of
Lemma~\ref{main:lemma:two}, and hence the proof of our main theorem,
Theorem~\ref{main:thm}.

Note that there is nothing special about the number 2 in our
quantifier elimination procedure: inspection of the proofs shows that
the arguments go through unchanged for any real algebraic number
$\alpha > 1$. There are various ways to represent the real algebraic
numbers; for example, we can represent $\alpha$ by providing a
polynomial, $p(x)$, of which it is a root, together by a pair of
rational numbers $u$ and $v$ isolating $\alpha$ from the other roots
of $p$. In that case, we simply replace $2$ by a new constant, $c$, in
the axioms, and then add the following:
\begin{itemize}
\item $p(c) = 0$
\item $u < c < v$
\end{itemize}
As noted in \cite{van:den:dries:gunaydin:unp}, this implies that the
resulting theory is decidable. To see this, it suffices to see that
any quantifier-free sentence $\ph$ is decidable. But we can do 
this
using the decision procedure for real closed fields to iteratively
compute the values of $\lambda(t)$ for any $t$ involving the field
operations and $c$, and then to determine the truth of terms of atomic
formulas $D_n(t)$. (For explicit algorithms for computing with real
algebraic numbers, see \cite{basu:et:al:03,mishra:93}.)


\section{Complexity analysis}
\label{complexity:section}

In this section we establish an upper bound on the complexity of our
elimination procedure.

For the theory of real closed fields, the best known upper bound for a
quantifier-elimination procedure, in terms of the length of the input
formula, is $2^{2^{O(n)}}$. This is originally due to Collins
\cite{collins:75}, and, independently, Monk and Solovay. There are
more precise bounds that depend on various parameters, such as the
number of quantifier alternations and the degrees of the polynomials
in the formula; see, for example, \cite{basu:99,basu:et:al:03}. In
particular, a block of existential quantifiers can be eliminated in
time $2^{O(n)}$. The best lower bound for the full
quantifier-elimination procedure is $2^{O(n)}$, by Fischer and Rabin
\cite{fischer:rabin:74}, and applies even to just the additive
fragment. The best upper bound for Presburger arithmetic is
$2_3^{O(n)}$ (see \cite{ferrante:rackoff:79,weispfenning:90}) and is
essentially sharp (see \cite{weispfenning:97}).

Our bounds are far worse. Consider what our procedure does when given
a formula with a single block of existential quantifiers:
\begin{enumerate}
\item First, replace this by a disjunction of formulas of the form
\[
\ex {\vec y} (A(\vec y) \land \ex {\vec z} (1 < \vec z < 2 \land
\psi))
\]
where $\psi$ is in the language of real closed fields.
\item Then, use an elimination procedure for real closed fields to
  eliminate the quantifiers $\ex {\vec z}$.
\item Successively eliminate the innermost quantifier over a power of
  two, as follows:
\begin{enumerate}
\item Call the relevant formula $\ex x (A(x) \land \ph)$. Apply
  Proposition~\ref{main:prop}, to reduce $\ph$ to a formula that is
  simple in $x$.
\item Put the new $\ph$ in disjunctive normal form, split across a
  disjunction, and remove atomic formulas that do not involve $x$, so
  that each formula is of the form
\[
\ex x (A(x) \land \bigwedge p_i(x) = 0 \land \bigwedge q_j(x) = 0 \land
\theta)
\]
where $\theta$ is a conjunction of formulas of the form $D_n(2^r x)$
and negations of such, and in each disjunction where a disjunct of the
form $p(x) = 0$ occurs, we can assume $p$ is not identically $0$.
\item In each disjunct where a conjunct of the form $p(x) = 0$ occurs,
  apply Lemma~\ref{algebra:two:lemma}, factor out the divisibility
  predicates, $D_n$, and call a quantifier-elimination procedure for
  real closed fields.
\item In the remaining disjuncts, again, split across a disjunct; in
  one case, we call a quantifier-elimination procedure for real closed
  fields right away; in another, we expand a bounded existential
  quantifier into a disjunction, and then call the elimination
  procedure for real closed fields.
\end{enumerate}
\end{enumerate}

Note that each iteration of the inner loop (3) requires at least one
call to a quantifier-elimination procedure for real closed fields.
Each of these calls can be carried out in time, say, $2^{2^{O(n)}}$,
where $n$ is the length of the relevant formula. But then the next
iteration of the loop will involve calls to the q.e.~procedure for
real closed fields on a formula that is potentially much longer. Thus,
part (3) of the procedure requires an exponential stack of $Cm$ twos,
for some constant $C$, where $m$ is the number of existential
quantifiers over powers of two that need to be eliminated.

In this section, we will confirm that such an upper bound can be
obtained. To that end, it is sufficient to show that each pass of the
inner loop is elementary, which is to say, it can be computed in time
bounded by some fixed stack of exponents to the base 2. Note that
after the first step, the number of quantifiers over powers of two is
bounded by the length of the original formula (in fact, it is bounded
by the number of $A$'s and $\lambda$'s in the original
formula). Thus our procedure for eliminating a block of existential
quantifiers runs in time $2_{O(n)}^0$, where $n$ is the length of the
original formula.

We have been unable to eliminate this nesting of calls to a procedure
for real closed fields. Efficient procedures for this latter theory
avoid putting formulas in disjunctive normal form; for example,
Collins's cylindrical algebraic decomposition procedure obtains a
description of cells, depending on the coefficients, on which a set of
polynomials have constant sign. In our setting, suppose we are given a
formula $\ex {\vec x} (A(\vec x) \land \eta \land \theta)$, where
$\eta$ contains only equalities and inequalities between polynomials,
and $\theta$ consists of divisibility conditions $D_n$ on the
exponents of the $x$'s. One might start by applying Collins's procedure
to the polynomials occurring in $\eta$. Then, given a description of
the various cells (depending on the other parameters in the formula),
one needs to determine which cells contain points with coordinates
that are powers of two, with exponents satisfying the requisite
divisibility conditions. For one dimensional cells, our procedure
relies on a simple disjunction: if the cell is large enough, one is
guaranteed a solution, and otherwise one need only test a finite
number of cases.  For multidimensional cells, however, the situation
is more complex, and we do not see how one can proceed except along
the lines we have described above. It is thus an interesting question
as to whether it is possible to obtain elementary bounds on a
procedure for eliminating a single block of quantifiers. Given our
failure to do so, we have not taken great pains to bound the number of
exponents in the time bound on the inner loop, which would merely
improve the constant bound implicit in the $O(n)$.

For the discussion which follows, we define the \emph{length} of a
formula in the language of $T$ to be the number of symbols in a
reasonable formulation of the first-order language, with the following
exception: we count the length of each symbol $D_n$ as $n$, rather
than, say, one plus the binary logarithm of $n$. This choice is a
pragmatic one in that it simplifies the analysis, and our results
below then imply the corresponding results for the alternative
definition of length. A more refined analysis might take both the
length of the formula and a bound on the $n$'s occurring in atomic
formulas $D_n(t)$, but that does not seem to help much.

It seems that the most delicate part of our task is showing that one
can remove the division symbols, and ``squeeze'' variables ranging
over powers of two out of the $\lambda$ symbols that are repeatedly
introduced after the first step of the procedure, as required in step
(3a). A priori, the procedures described in
Section~\ref{reasoning:section} look as though they may be
non-elementary. The next few lemmas show that this is not the case, by
keeping careful track of the terms and formulas that need to be dealt
with in the disjunctions.

\begin{lemma}
\label{messy:zero}
  Let $t$ be a term with length $l$. Then there is a sequence of terms
  $\la t_k: k < 2^l \ra$ such that
\begin{itemize}
\item $T \proves \bigvee_{k < 2^l} t = t_k$,
 \item each $t_k$ is of the form $r / s$, where $r$ and $s$ are
 division-free terms, and
\item each $t_k$ has length at most $2^l$.
\end{itemize}
\end{lemma}

\begin{proof}
  This can be proved by a straightforward induction on terms. Suppose
  $t$ is of the form $t_1 + t_2$, where the length of $t_1$ is $l_1$
  and the length of $t_2$ is $l_2$. By the inductive hypothesis, $t$
  is equal to one of at most $2^{l_1} 2^{l_2} \leq 2^l$ terms of the
  form $r_1 / s_1 + r_2 / s_2$, where $r_1$, $s_1$, $r_2$, and $s_2$
  are division-free, the length of $r_1 / s_1$ is at most $2^{l_1}$,
  and the length of $r_2 / s_2$ is at most $2^{l_2}$. But then the
  length of $(r_1 s_2 + r_2 s_1) / s_1 s_2$ is at most $2 (2^{l_1} +
  2^{l_2}) < 2^l$, as required.
  
  If $t$ is of the form $\lambda(t_1)$, the claim follows from the
  inductive hypothesis, using Lemma~\ref{elim:div:lambda:lemma}. The
  other cases are similar.
\end{proof}

\begin{lemma}
\label{messy:one}
Let $\varphi$ be a quantifier-free formula with length $l$. Then there
is a quantifier-free division-free formula $\varphi'$ with length
$2^{O(l)}$ such that $T \proves \varphi \leftrightarrow \varphi'$.
\end{lemma}

\begin{proof}
Enumerate all the different terms $t_0, \ldots, t_{m-1}$ in
$\varphi$ such that, for each $i < m$, $s_i$ is not a proper subterm
of any term in $\varphi$. Using the above lemma we can have a
sequence of quantifier-free formulas $\varphi_j$ for $j < 2^l$ each
of which is obtained by replacing each $t_i$ with an appropriate
term and therefore has length less than $2^l$. Notice that
for each $\varphi_j$, as indicated in
Lemma~\ref{elim:div:lambda:lemma}, there are some division-free
atomic formulas that $T$ used to derive the equalities in question.
Clearly for each $\varphi_j$ there are less than $l$ such atomic
formulas, each of which has length less than $2^{O(l)}$. Let
$\sigma_j$ be the conjunction of them all. Let $\varphi'$ be the
formula $\bigvee_{j < 2^l} (\varphi_j \wedge \sigma_j)$. The length
of $\varphi'$ is again bounded by $2^{O(l)}$, and clearly $T
\proves \varphi \leftrightarrow \varphi'$.

Finally, we need to clear denominators from atomic formulas of the
form $r / s < t / u$ and $r / s = t / u$, and deal with atomic
formulas of the form $D_n(r/s)$. The first two require a disjunction
over cases, depending on whether denominators are positive, negative,
or zero. The third set of atomic formulas is handled as described in
the proofs of Lemma~\ref{elim:one:div:prop},~\ref{elim:div:prop}. But
each atomic formula occurring in a disjunct occurs to an atomic formula
in the original formula, $\ph$, and there are at most $l$ of these. It
is not hard to verify that the corresponding increase in length can be
absorbed into the bound $2^{O(l)}$.
\end{proof}

\begin{lemma}
\label{messy:two}
Let $\lambda(t)$ be a term, where the length of $t$ is $l$ and $x$
does not occur in the scope of any division symbol in $t$. Then there
is a sequence of terms $\langle t_k: k < 2^{8l^2\log l} \rangle$ such
that
\begin{itemize}
 \item $T \proves A(x) \land t > 0 \limplies \bigvee_{k < 2^{8l^2\log l}} (\lambda(t) = t_k)$,
 \item each $t_k$ is of the form $sx^i$, where $s$ is a term
 that does not contain $x$ and $i < l$,
 \item each $t_k$ has length at most $2^{2^{4l}}$.
\end{itemize}
\end{lemma}

\begin{proof}
For any polynomial $p$ in $x$, clearly the number of possible values
of $\lambda(p)$ of the form $sx^i$, as in
Lemma~\ref{algebra:three:lemma}, depends on the degree $n$ of $x$ in
$p$. So let $f(n)$ denote the number of possible values of
$\lambda(p)$. Observe that the value of $\lambda(p)$ is determined in
the first case of Lemma~\ref{algebra:one:lemma}, and when $e = 1$ in
the second case. An calculation shows that there are no more than
$(n+1)(n+2)$ possibilities in the first case, no more than $2n(2n+2)$
possibilities in the second case when $e = 1$, and no more than
$(n+1)(n-1)2(n+2)$ possibilities for all the remaining values of
$e$. Hence we have the following equation:
\[
f(n) \leq (n+1)(n+2) + 2n(2n+2) + (n+1)(n-1)2(n+2)f(n-1).
\]
This can be simplified as $f(n) < 10(n+2)^3f(n-1)$. So we have $f(n)
< 2^{8n\log(n+2)}$. Let the length of $p$ be $l$. Since $n+2 < l$,
we have $f(n) < 2^{8l\log l} < 2^{8l^2\log l}$.

Now the proof proceeds by induction on the $\lambda$-depth of $x$
in $t$. If $\Lambda(x, t) = 0$, then $t$ is a polynomial in $x$. So
we apply the above analysis to $t$ and obtain no more than
$2^{8l\log l}$ possible values of $\lambda(t)$ which are all of the
form $sx^i$ for some $i < l$. To compute the length of $s$, only
note that each step of the iteration produces a polynomial whose
length is no more than the square of the length of the previous
polynomial. So we conclude that the length of $s$ is no more than
$l^{2^l} < 2^{2^{4l}}$.

Now suppose the lemma holds for each term $s$ with $\Lambda(x, s) <
d$, and suppose $\Lambda(x, t) = d$. Enumerate all the different terms
$\lambda(s_0), \ldots, \lambda(s_{m-1})$ in $t$ such that
$\lambda(s_i)$ is not in the scope of any $\lambda$ for each $i < m$.
Clearly $\Lambda(x, s_i) < n$ for each $i < m$. So by the inductive
hypothesis there are less than $2^{8l_i^2\log l_i}$ possible values
for each $\lambda(s_i)$, where $l_i$ is the length of $s_i$. Since
$\sum_{i < m} l_i < l-1$, there are no more than $2^{8(l-1)^2\log l}$
possible values for $t$. Enumerate these possibilities as $\langle
t_k: k < 2^{8(l-1)^2\log l} \rangle$. In each $t_k$, $\lambda(s_i)$ is
replaced by a term of the form $sx^j$ with $j < l_i$. So $t_k$ is a
polynomial in $x$ whose degree in $x$ is less than $l-2$. So there are
$2^{8(l-1)^2\log l} \cdot 2^{8l\log l} \leq 2^{8l^2\log l}$ possible
values for $\lambda(t)$. The length of each $t_k$ is bounded by
$2^{2^{4(l-1)}}$, so the length of each possible value of $\lambda(t)$
is bounded by $2^{2^{4(l-1)}} \cdot l \cdot l^{2^l} < 2^{2^{4l}}$.
\end{proof}

\begin{lemma}
\label{messy:three}
Let $\varphi$ be a quantifier-free formula with length $l$. Assume $x$
does not occur in the scope of any division symbol in $\varphi$. Then
there is a quantifier-free formula $\varphi'$ with length at most
$2^{2^{O(l)}}$ such that $\varphi'$ is simple in $x$ and $T \proves
A(x) \limplies (\varphi \leftrightarrow \varphi')$.
\end{lemma}

\begin{proof}
First we claim there is a quantifier-free formula $\varphi^*$ with
length at most $2^{2^{O(l)}}$ such that
\begin{itemize}
 \item $T \proves A(x) \limplies (\varphi \leftrightarrow
 \varphi^*)$,
 \item $x$ does not occur in the scope of any division in $\varphi^*$,
 \item $\Lambda(x, \varphi^*) = 0$.
\end{itemize}
The proof is essentially the same as the proof of
Lemma~\ref{messy:one}, using Lemma~\ref{messy:two} instead of
Lemma~\ref{messy:zero}.

Next we need to deal with atomic formulas of the form $D_n(p)$ in
$\varphi^*$, as shown in Lemma~\ref{algebra:five:lemma}. So $p$ is a
polynomial in $x$ whose degree in $x$ is less than $l$. So there are
at most $2^{2^{O(l)}}$ possible values for $\lambda(p)$, the length
of each of which is bounded by $2^{2^{O(l)}}$. So each $D_n(p)$ can
be replaced by a disjunction whose length is less than
$2^{2^{O(l)}}$. So the bound does not change.

The increase in length in transforming $\varphi^*$ to a formula that
is simple in $x$, as described in the proof of
Lemma~\ref{algebra:six:lemma}, can be absorbed in the bound
$2^{2^{O(l)}}$.
\end{proof}

\begin{lemma}
\label{messy:last}
Let $\varphi$ be a quantifier-free formula with length $l$. Then
there is a quantifier-free formula $\varphi'$ with length at most
$2^{O(l)}_3$ such that $\varphi'$ is simple in $x$ and $T$ proves
$A(x) \limplies (\varphi \leftrightarrow \varphi')$.
\end{lemma}

\begin{proof}
Immediate by Lemma~\ref{messy:one} and Lemma~\ref{messy:three}.
\end{proof}

\begin{lemma}
Each iteration of step 3 can be performed by an elementary function. 
\end{lemma}

\begin{proof}
  It is straightforward to verify that the procedure implicit in
  Lemmas~\ref{messy:last} runs in time polynomial in its output. As a
  result, step 3(a) is elementary.  Step 3(b) is also clearly
  elementary. In fact, even though putting a formula in disjunctive
  normal form can result in exponentially many disjuncts, since each
  disjunct only involves atomic formulas from the original formula,
  the length of each disjunct is bounded in the length of the original
  formula. 
  
  After step 3(a), the main increase therefore comes from the handling
  of the cases in (c) and (d), each of which is easily seen to be
  elementary. Case (c) involves a call to a quantifier-elimination
  procedure for real closed fields, with a $\forall \exists$ formula;
  case (d) involves calls to such a procedure, on existential
  formulas, across a number of disjuncts that is exponential in
  the length of the original formula.
\end{proof}

\begin{theorem}
  There is a procedure for eliminating a single block of existential
  quantifiers in theory $T$ in time $2_{O(l)}^0$, where $l$ is the
  length of the original formula.
\end{theorem}

\begin{proof}
Steps 1 and 2 are clearly elementary, after which the procedure
performs an elementary operation for each quantifier over a power of
two. As noted above, the number of such quantifiers can even be
bounded by the number of predicates $D_n$ and $\lambda$'s in the
original formula.
\end{proof}

\begin{corollary}
There is a procedure for eliminating quantifiers in theory $T$ that
runs in time bounded by $O(l)$ iterations of the stack-of-twos
function, where $l$ is the length of the original formula.
\end{corollary}

\begin{proof}
Put the formula in prenex form, and iteratively apply the previous
theorem to eliminate each block of quantifiers.
\end{proof}


\nocite{caviness:johnson:93}

\bibliographystyle{plain} 
\bibliography{proofthry}

\end{document}